\documentclass{ws-procs975x65}

\begin{document}

\title{MODEL OF A DISK-OUTFLOW COUPLED SYSTEM: DISK-OUTFLOW SYMBIOSIS} 

\author{BANIBRATA MUKHOPADHYAY}

\address{Department of Physics, Indian Institute of Science,
Bangalore 560012\\
E-mail: bm@physics.iisc.ernet.in\\
}

\begin{abstract}
Several observational evidences and deeper theoretical insights reveal that accretion
and outflow/jet are strongly correlated. We model an advective disk-outflow coupled
dynamics. 
We investigate the
properties of the disk-outflow surface and how is it dependent on the spin
of the black hole. The energetics of such a symbiotic system strongly depend on
the mass, accretion rate and spin of the black holes. The model clearly shows that
the outflow power extracted from the disk increases strongly with the spin of the
black hole.
\end{abstract}


\bodymatter

\section{Introduction}\label{intro}

There are many observational evidences for strong outflows and
jets in black hole accreting systems, both in 
microquasars (e.g. SS433, GRS 1915+105)\cite{mirabel} and
quasars or active galactic nuclei (AGNs) (e.g. Sgr~$A^*$)\cite{begel}.
Extragalactic radio sources show evidences for strong jets\cite{meier}. 
As the jets
are coming from the disk, base of the jets can not be treated 
independent of disk, which
might be influenced by the spin of black holes. 
Outflows/jets take matter out of the 
disk, which helps in removing angular momentum of the disk and
infalling the matter. Hence, inflow/disk and outflow/jet
governing from a same system can not be treated independently. Indeed,
observed data argue for the correlation of radio to X-ray emissions\cite{fender}.

Most of the jet emissions are associated with the low/hard state of 
accretion disk (e.g. GRS~1915+105). On the other hand, a low/hard state 
corresponds to a non-Keplerian flow\cite{sle}.
Hence, here we report the coupled disk-outflow dynamics in the framework
of a sub-Keplerian, advective, geometrically thick accretion disk. 
For a detailed description, see Ref.~\refcite{bgm}.

\section{Basic equations}

We consider the correlated dynamics, governed by conservation laws,
based on bi-directional hypothesis: outflow/jet is necessary to satisfy 
disk boundary condition(s). Hence, we plan to solve the set of coupled partial differential
hydrodynamic equations in the advective regime, when 
flow variables depend on radial ($r$) and vertical ($z$) coordinates. 
However, we neglect the effects
of magnetic field, as we do not aspire to describe the jet mechanism.
Moreover, there are models (CENBOL\cite{chak99}, ADAF\cite{ny}), arguing for
outflows without magnetic effects. 
Indeed, supercritical accretions predict outflows/jets
by strong radiation pressure\cite{abrapir}.
In addition, we neglect viscosity as we concentrate on a predefined inner region of
the system where outflowing matter is able to remove angular momentum and then transport matter.
Hence, the hydrodynamic balance equations along with the equation of continuity in the 
pseudo-Newtonian framework are give by
\begin{eqnarray}
\nonumber
&&\frac{1}{r} \frac{\partial}{\partial r} (r \rho v_r) +  \frac{\rho v_z}{z}=  0,\,\,
v_r \frac{\partial v_r}{\partial r} +  v_z \frac{v_r-v_{r0}}{z}  -  \frac{\lambda^{2}}{r^3}  +  F_{Gr}  +  \frac{1}{\rho} \frac{\partial P}{\partial r}   =  0,\\
&&v_r \frac{\partial v_z}{\partial r}  +  \frac{v^{2}_z}{z}  +   F_{Gz}  +  \frac{1}{\rho} \frac{P-P_0}{z}  =  0,\,\,
\frac{d\lambda}{dt}=0,
\label{5b}
\end{eqnarray}
where we choose $\partial/\partial z\equiv 1/z$ as the radial variations of the 
underlying variables are much stronger than their vertical variations so that the flow structure
remains disk. Here $v_r$, $v_z$, $\lambda$, $\rho$, $P$ define respectively radial, vertical
velocities, specific angular momentum, density, pressure of the flow and those with subscript `$0$' are the
respective variables at $z=0$. The above equations are to be solved with the 
condition $P|_h=0$, when $h$ is the disk scale height, applying in the vertical momentum balance
equation at $z=h$ 
and by the prescription of two-dimensional pseudo-Newtonian potential given by Ref.~\refcite{gm}.
We also assume $P=K\rho^\gamma$, where $K$ is a constant and $\gamma$ the adiabatic index, and
the sound speed $c_s=\sqrt{\gamma P/\rho}$. Presuming that the outflow velocity is not likely to 
exceed the sound speed at the disk-outflow surface, we
further propose $v_z={\it l}(z/r)^\mu\,c_s$, when $\it l$ and $\mu$ are the constant parameters. 

\section{Solutions}

\begin{figure}[]%
\begin{center}
\hskip-1.5cm
 \parbox{2.0in}{\epsfig{figure=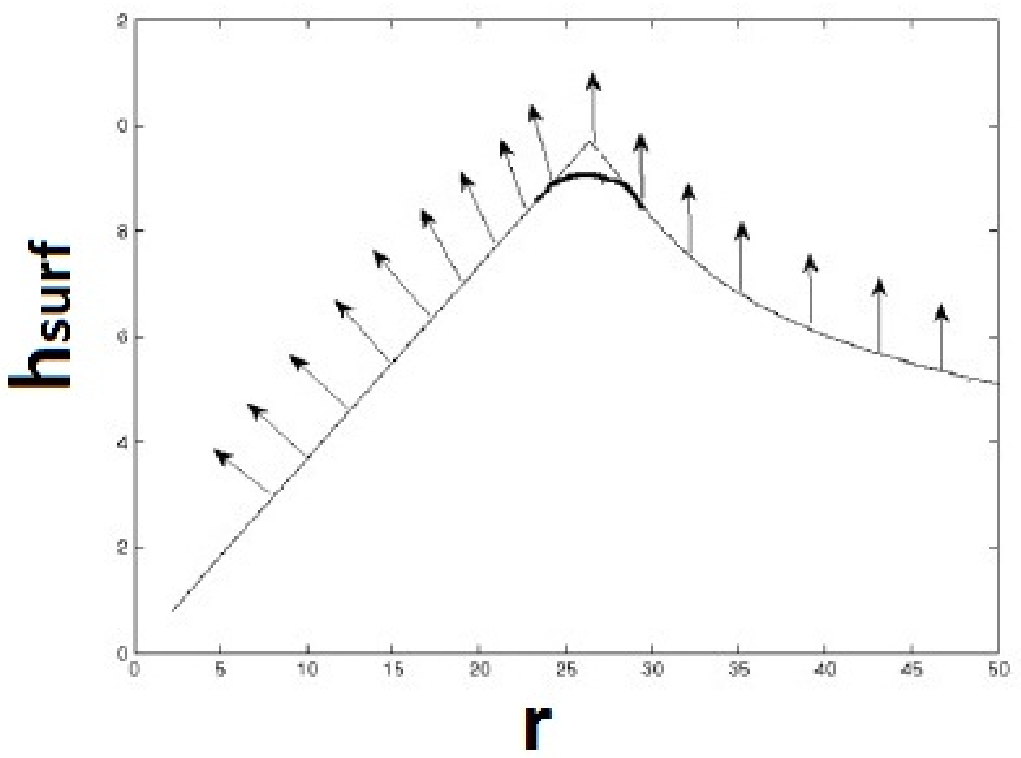,width=2.2in}
}
 \hspace*{4pt}
 \parbox{2.1in}{\epsfig{figure=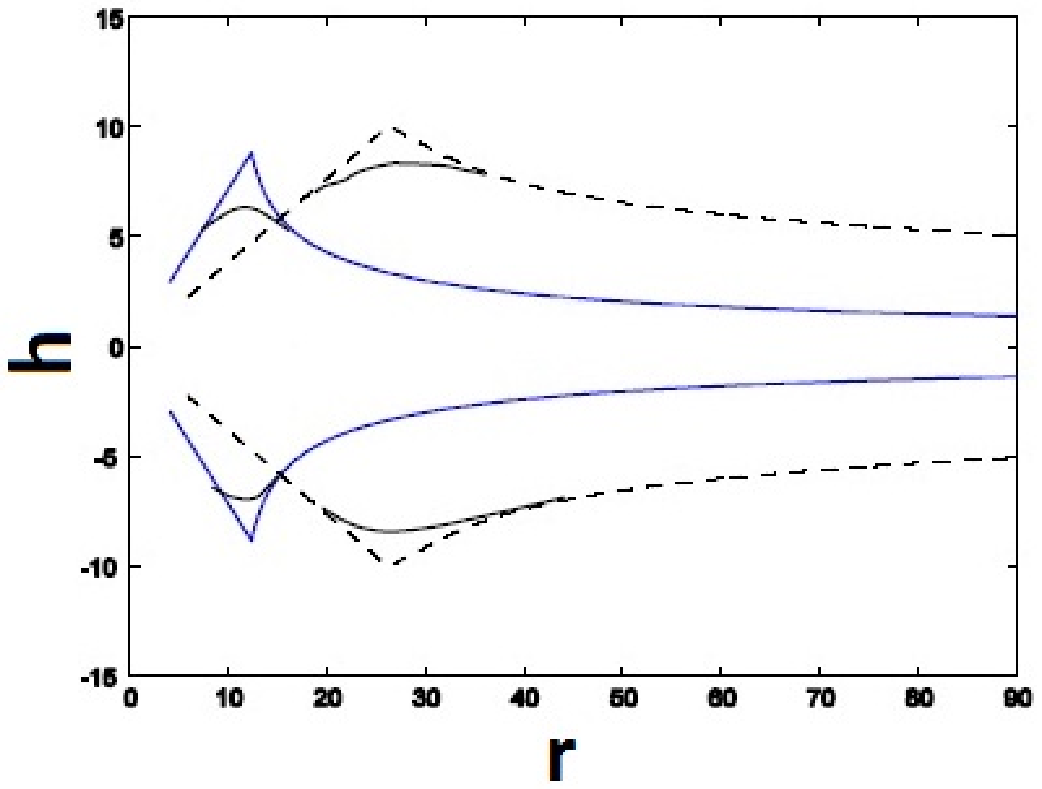,width=2.4in}
}
\vskip0.5cm
\hskip-.5cm
 \parbox{2.3in}{\epsfig{figure=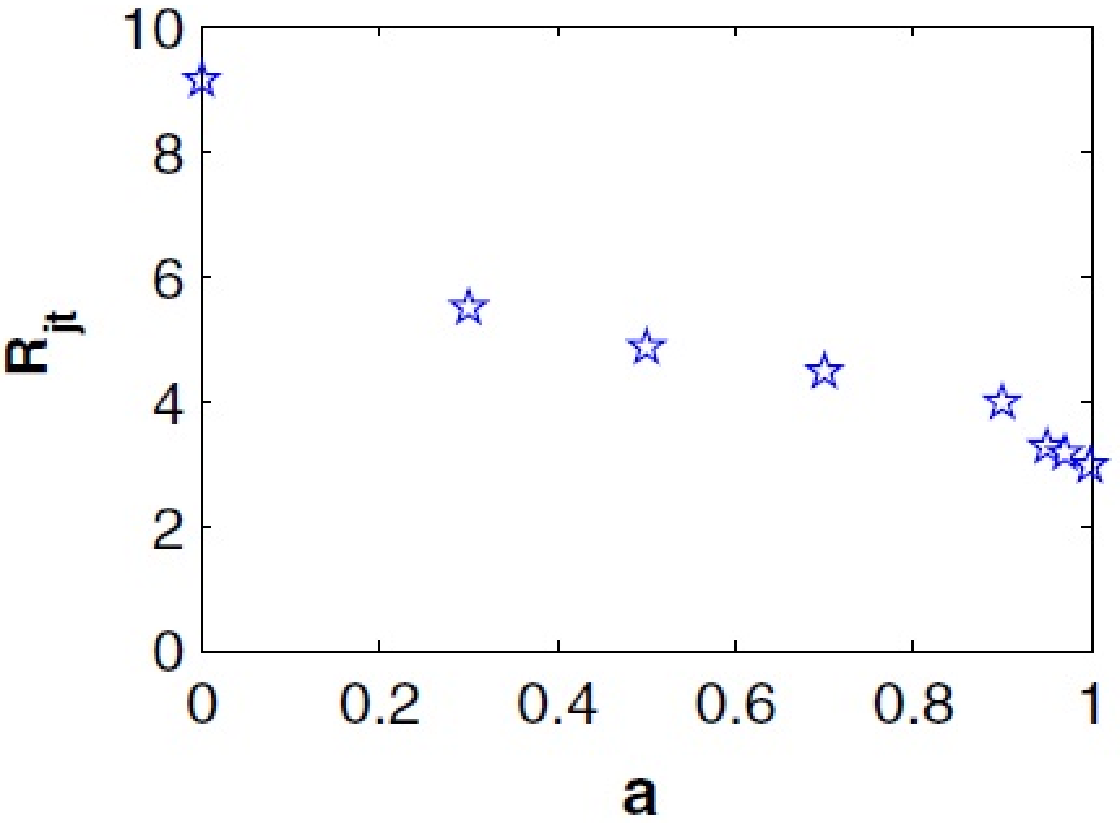,width=2.0in}
}
 \parbox{2.3in}{\epsfig{figure=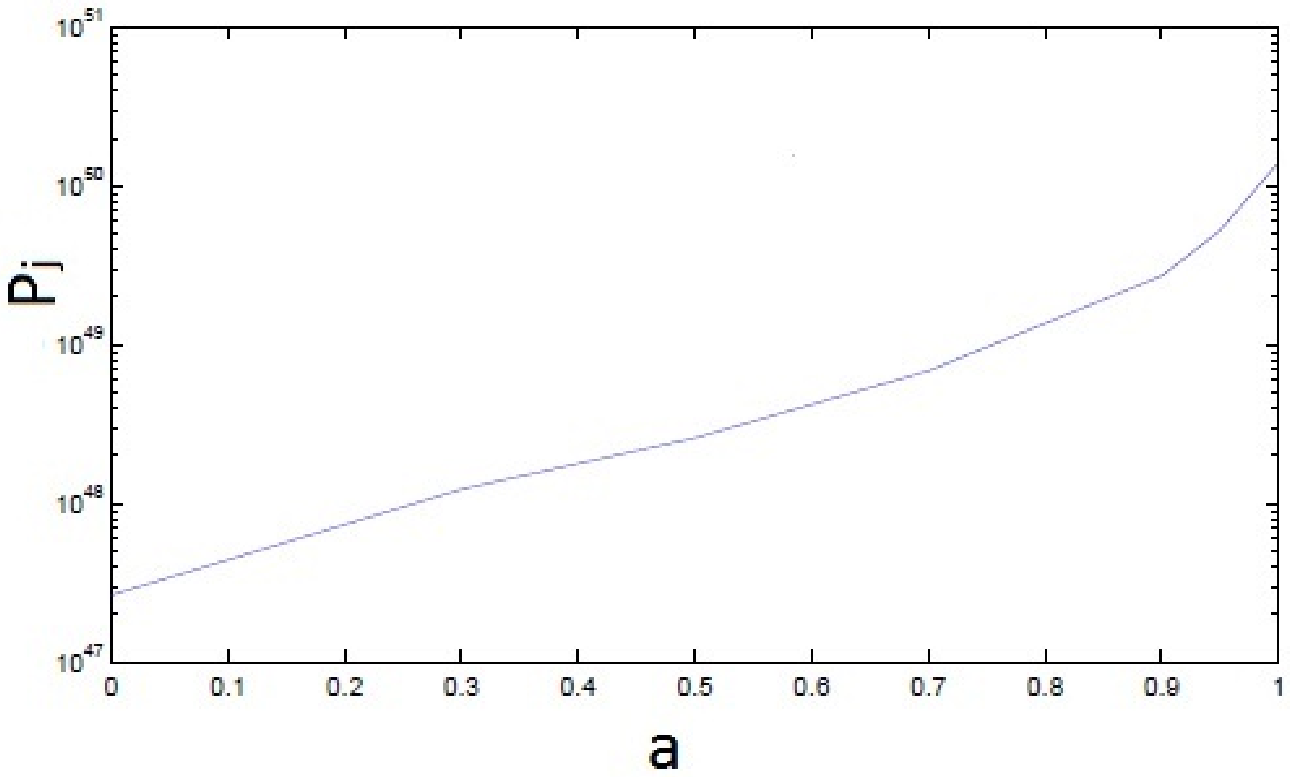,width=2.4in}
}
 \caption{Upper-left (a): Nature of the surface of disk-outflow coupled region.
Upper-right (b): Geometry of disk-outflow coupled
region, solid and dashed lines are respectively for $a=0.998$ and $0$.
Lower-left (c): Variation of inner radius ceasing disk-outflow coupled region
as a function of
black hole's spin. Lower-right (d): Outflow power in C.G.S. unit as a function of black hole's spin,
with one Eddington accretion rate around a $10^7$ solar mass black hole.
}
\label{fig}
\end{center}
\end{figure}

Solving the set of eqns. (\ref{5b}), we find
the disk-outflow coupled region ceased to exist at a $z=h$ where $v_r=0$, as outflow takes 
away the matter. 
Figure \ref{fig}a shows the upper surface of
the coupled region ($h_{\rm surf}$). The arrows in the diagram
reveal the direction of flow. 
A more realistic disk-outflow surface could
be visualized with a thick-solid line around the peak shown in Fig \ref{fig}a. 
Figure \ref{fig}b shows the geometry of the entire disk-outflow coupled region and 
that how the region changes with the change of spin of the black hole ($a$). Larger the spin,
thinner the region is, and advancing the peak of the region towards the black hole
(i.e. the system becomes more prone to produce outflows in the vicinity of black hole).
This is because, a faster spinning black hole is more prone to eject matter
out leaving a smaller remnant. Additionally, stronger centrifugal effect of a faster
spinning black hole ejects matter from a closer vicinity of the black hole compared to
a slower black hole. Figure \ref{fig}c shows the variation of the radius ceasing the
disk-outflow coupled region in the vicinity of the black hole ($R_{jt}$) 
with $a$, reflecting the same feature mentioned above. Finally, Fig. \ref{fig}d shows clearly 
the increase of outflow (and hence jet) power
with increasing $a$. Note that the mechanical outflow power is defined as
$
P_j (r) = \int 4 \pi r \biggl[\biggl(\frac{v^2}{2} + \frac{\gamma}{\gamma-1} 
\frac{P}{\rho} + \phi_G  \biggr) \rho v_z \biggr] \bigg |_{h_{surf}} \, dr, 
$
where $v$ is the magnitude of total velocity and $\phi_G$ the gravitational potential.

\section{Conclusions}

It is found that the power of astrophysical jets increases
with the increasing spin of central object. In case of blazars,
emissions are believed to be originated
from their jets, whose high energy properties can be understood from this theory;
see Ref.~\refcite{mbs}.
If the extreme gravity is responsible for powering strong outflows and jets,
then the spin of the black hole, perhaps, is the fundamental parameter to
account for the observed astrophysical processes. \\

This work was partially supported by the grant ISRO/RES/2/367/10-11.







\end{document}